\documentclass[namedreferences,hyperref,optionalrh,solaromanenum]{texsupport.springer-sola-master/spr-sola}
\usepackage{amsmath,amssymb}
\usepackage{lineno}
\usepackage{graphicx}
\usepackage{color}
\usepackage{booktabs}
\usepackage{float}
\usepackage{url}

\newcommand{\apj}{{\it Astrophys. J.}}

\newcommand{\jgr}{{\it J. Geophys. Res.}}

\newcommand{\solphys}{{\it Sol. Phys.}}
\newcommand{\ssr}{{\it Space Sci. Rev.}}
\chardef\us=`\_

\newcommand{\etal}{et~al.}

\begin{document}

\begin{frontmatter}

\title{LAGO Detection of the January 2026 Forbush Decrease in Antarctica and Near the Tropic of Capricorn}

\author[addressref={aff7}]{%
  \inits{Y.D.}\fnm{Yessica~Dominguez}~\snm{Ballesteros}\orcid{0009-0007-0462-9630}}
\author[addressref={aff2}]{%
  \inits{J.F.S.}\fnm{João~Francisco~Silva}~\snm{Correia}\orcid{0009-0008-7420-220X}}
\author[addressref={aff3,aff4}]{%
  \inits{S.}\fnm{Sergio}~\snm{Dasso}\orcid{0000-0002-7680-4721}}
\author[addressref={aff8}]{%
  \inits{M.S.}\fnm{Mauricio Suárez}~\snm{Durán}\orcid{0000-0003-1451-630X}}
\author[addressref={aff2},corref,email={fauth@unicamp.br}]{%
  \inits{A.C.}\fnm{Anderson}~\snm{Campos~Fauth}\orcid{0000-0001-7239-0288}}
\author[addressref={aff5}]{%
  \inits{G.B.S.}\fnm{Giovanni~Battista~Secchia}~\snm{González}\orcid{0009-0004-3420-0497}}
\author[addressref={aff3,aff4}]{%
  \inits{A.M.}\fnm{Adriana~Maria}~\snm{Gulisano}\orcid{0000-0002-2234-4432}}
\author[addressref={aff5}]{%
  \inits{J.}\fnm{Jorge}~\snm{Molina}\orcid{0000-0002-5885-0206}}
\author[addressref={aff5,aff6}]{%
  \inits{J.}\fnm{Jesus}~\snm{Nuñez}\orcid{0000-0002-2842-1582}}
\author[addressref={aff7}]{%
  \inits{L.A.}\fnm{Luis~A.}~\snm{Núñez}\orcid{0000-0003-4575-5899}}  
\author[addressref={aff1}]{%
  \inits{N.A.}\fnm{Noelia~Ayelén}~\snm{Santos}\orcid{0000-0003-2588-1054}}
\author[addressref={aff9}]{%
  \inits{C.}\fnm{Christian}~\snm{Sarmiento-Cano}\orcid{0000-0002-4977-4184}}
\author[addressref={aff10}]{%
  \fnm{On~behalf~of~the}~\snm{LAGO~Collaboration}}

\address[id=aff1]{Instituto Tecnológico de Buenos Aires, Buenos Aires, Argentina}
\address[id=aff2]{University of Campinas, Campinas, Brazil}
\address[id=aff3]{Instituto de Astronomía y Física del Espacio, Buenos Aires, Argentina}
\address[id=aff4]{Universidad de Buenos Aires, Buenos Aires, Argentina}
\address[id=aff7]{Universidad Industrial de Santander, Bucaramanga, Colombia}
\address[id=aff5]{Universidad Nacional de Asunción, Asunción, Paraguay}
\address[id=aff6]{National Institute for Space Research, São José dos Campos, Brazil}
\address[id=aff8]{Universidad de la Costa, Barranquilla, Colombia}
\address[id=aff9]{Universidad Autónoma de Bucaramanga, Bucaramanga, Colombia}
\address[id=aff10]{https://lagoproject.net/}

\runningauthor{Dominguez Ballesteros \etal}
\runningtitle{Forbush Decreases with {LAGO}}

\begin{abstract}
We report the detection of the 19\,--\,20 January 2026 Forbush decrease (FD)
with four ground-level detectors of the Latin American Giant Observatory
(LAGO): water-Cherenkov detectors (WCDs) at the Antarctic stations of Marambio
and San Mart\'{\i}n (vertical geomagnetic cutoff rigidity
$R_c \approx 1.6$\,--\,$2.1$~GV) and at TANCA (Campinas, Brazil;
$R_c \approx 9.3$~GV), and a scintillator coincidence detector at Asunci\'on
(Paraguay; $R_c \approx 10.7$~GV). The event was driven by a full-halo coronal
mass ejection (CME) associated with an X1.95 flare from active region AR~4341
on 18 January. We reconstruct its early kinematics with the Graduated
Cylindrical Shell model and propagate it to 1~AU with the Drag-Based Model;
the fast ($\approx 25$~h) transit requires a low drag parameter,
consistent with an interplanetary medium preconditioned by
preceding CMEs. The interplanetary shock reached L1 at 19:17~UT on
19 January, and the FD began nearly simultaneously at all four stations at
about 21:40~UT. All four stations recorded maximum depressions of 9.09\,--\,9.41\% despite their different local cutoff rigidities, whereas the Mexico neutron monitor ($R_c = 8.2$~GV) recorded a deeper ($\sim 17\%$) and slower-recovering decrease. The two-step decrease follows the passage of the
shock sheath and of the ICME ejecta. Simulations with the
Atmospheric Ray Tracing and Interaction framework reproduce the separation
between the electromagnetic and muonic components of the WCD response. These
multi-site, multi-technique observations illustrate the complementary
sensitivity of muonic and nucleonic detectors to cosmic-ray modulation.
\end{abstract}

\begin{keyword}
\kwd{Cosmic Rays, Galactic}
\kwd{Coronal Mass Ejections, Interplanetary}
\kwd{Magnetosphere, Geomagnetic Disturbances}
\kwd{Solar Wind, Disturbances}
\end{keyword}

\end{frontmatter}

\section{Introduction}
\label{S-intro}

Forbush decreases (FDs) are transient, global reductions in the
galactic cosmic-ray (GCR) flux at ground level caused by the passage of interplanetary coronal mass ejections (ICMEs) or co-rotating interaction regions \citep{Forbush1937,Cane2000}. The magnitude and temporal profile of an FD carry information about the magnetic topology of the perturbing solar-wind structure and are therefore key diagnostics
in space-weather research.

The Latin American Giant Observatory \citep{Asorey2018lago, SarmientoCanoEtal2026} is a distributed network of single-particle-counting water-Cherenkov detectors (WCDs) spanning a wide range of vertical geomagnetic cutoff rigidities ($R_c$): from below 2~GV at high-latitude Antarctic stations to values exceeding 10~GV at tropical and equatorial sites. This broad $R_c$ coverage allows FDs to be observed simultaneously at sites with very different geomagnetic shielding.

In this work, we analyse the Forbush decrease of 19--20 January 2026, associated with the CME launched on 18 January 2026 (Sect.~\ref{S-cme}), using four
LAGO stations:
\begin{itemize}
  \item Marambio and San~Mart\'in (Antarctic Peninsula, $R_c \approx 1.6$--$2.1$~GV), described in Sect.~\ref{S-marambio} and~\ref{S-sanmartin};
  \item TANCA (Campinas, Brazil, $R_c \approx 9.3$~GV), Sect.~\ref{S-tanca};
  \item Asunci\'on (Paraguay, $R_c \approx 10.7$~GV), Sect.~\ref{S-asuncion}.
\end{itemize}
Section~\ref{S-cme} describes the solar origin of the event and the \textit{in-situ} solar-wind observations.
Section~\ref{S-detectors} describes the four detectors, and Section~\ref{S-arti} presents their simulated response to secondary radiation at ground level.
Section~\ref{S-results} presents the cosmic-ray observations, compares them with the Mexico neutron monitor and interplanetary data, and traces the CME propagation from the corona to 1~AU (Section~\ref{S-dbm}).
Section~\ref{S-conclusions} summarises our conclusions.

\section{Solar and Interplanetary Context}
\label{S-cme}

\subsection{Solar Origin}
\label{S-solarorigin}

\begin{figure}[t]
\centering
\begin{minipage}[b]{0.48\textwidth}
  \centering
  \includegraphics[width=\textwidth]{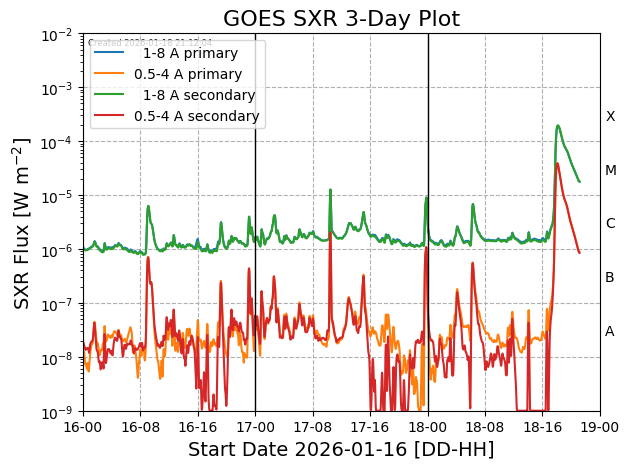}\\[2pt]
  {\small (a)}
\end{minipage}\hfill
\begin{minipage}[b]{0.48\textwidth}
  \centering
  \includegraphics[width=\textwidth]{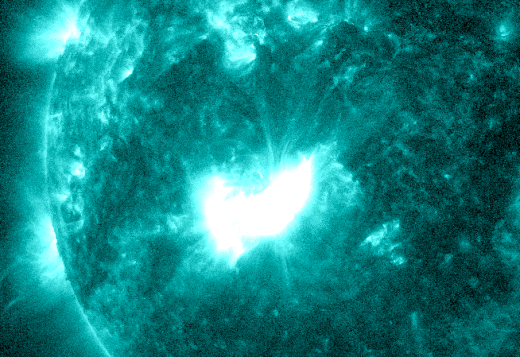}\\[2pt]
  {\small (b)}
\end{minipage}
\caption{Solar signatures of the 18 January 2026 X1.95 flare from AR~4341.
(a) GOES soft X-ray flux in the 1--8~\AA\ and 0.5--4~\AA\ channels over
16--19 January 2026, showing the impulsive rise to X-class levels at
18:09~UT on 18 January. (b) Extreme-ultraviolet image of the flaring region
at the time of peak emission, showing the flare ribbons over AR~4341.
Sources: (a) NOAA/GOES; (b) SDO/AIA 131~\AA.}
\label{F-cme}
\end{figure}

The eruption on 18 January 2026 originated in active region AR~4341 near the central meridian and was associated with a GOES X1.95 flare that peaked in soft X-rays at 18:09~UT (Figure~\ref{F-cme}). Figure~\ref{F-cme}a shows the GOES soft X-ray flux in the 1--8~\AA\ and 0.5--4~\AA\ channels over 16--19 January 2026, with an impulsive rise to X-class levels marking the flare. Figure~\ref{F-cme}b shows an extreme-ultraviolet image of the flaring region at peak emission, where bright kernels over AR~4341 trace the flare ribbons at the reconnection site. A full-halo CME was observed; although the bulk of the ejecta propagated eastward, its halo morphology indicates a substantial Earth-directed component, thus favouring high geoeffectiveness. The disturbance reached Earth's environment on 19 January, initiating a causal chain linking the interplanetary shock to the magnetospheric response and cosmic-ray modulation.

\subsection{\textit{In-Situ} Solar-Wind Observations}
\label{S-insitu}
From 18 to 24 January 2026, the near-Earth interplanetary medium evolved from relatively quiet solar wind conditions into one of the most disturbed intervals of Solar Cycle 25 following the arrival of an Earth-directed coronal mass ejection (CME). On 18 January, the upstream solar wind exhibited typical background characteristics, including low plasma density, moderate interplanetary magnetic field (IMF) intensity, and small oscillations of the Bz component around zero. The disturbance reached the Sun–Earth L1 point at approximately 19:17 UT on 19 January, where the CME-driven forward shock produced an abrupt increase in plasma density, magnetic field strength, and dynamic pressure. The total magnetic field intensified from about 6 to 86 nT within a few minutes, while the proton density increased nearly tenfold, marking the onset of a strongly compressed, turbulent sheath region. Because the concurrent solar energetic particle event contaminated ACE plasma measurements, reliable velocity measurements resumed later, indicating solar-wind speeds approaching 1000~km\,s$^{-1}$, consistent with the passage of a fast ICME. The sheath region exhibited large-amplitude fluctuations of the IMF Bz component, producing alternating intervals of southward and northward magnetic fields and facilitating efficient solar wind–magnetosphere coupling. About 21 minutes after the L1 shock was detected, the disturbance impacted Earth's magnetosphere, triggering a sudden storm and rapidly developing severe geomagnetic conditions that reached G4 on the NOAA scale at 19:38 UT. During 20–22 January, the spacecraft traversed the ICME ejecta (magnetic cloud), where the magnetic field remained substantially enhanced while Bz continued to rotate between northward and southward orientations, maintaining elevated geomagnetic activity. As the ejecta moved beyond Earth on 23–24 January, plasma density, magnetic field intensity, and IMF orientation gradually relaxed toward typical ambient solar-wind values, marking the recovery of the interplanetary environment.

\section{Experimental Description}
\label{S-detectors}
Three of the LAGO detectors used here are single-station WCDs recording the integrated flux of secondary
particles (mostly muons, electrons, and photons) generated by cosmic-ray interactions in the Earth's atmosphere. The Asunción station, however, uses a coincidence-based detector consisting of plastic scintillators read out by silicon photomultipliers (SiPMs). Table~\ref{T-detectors} summarizes the main characteristics of each station.
\begin{table}
    \caption{Main characteristics of the four LAGO stations.}
    \label{T-detectors}
    \begin{tabular}{lccccc}
        \hline
        \textbf{Station} & \textbf{Location} &
        \textbf{Lat (°)} & \textbf{Lon (°)} &
        \textbf{Alt (m)} & $R_c$ \textbf{(GV)} \\
        \hline
        Marambio   & Antarctica   & $-64.2$ & $-56.7$ & 198 & $\approx 2.1$ \\
        San~Mart\'in & Antarctica & $-68.1$ & $-67.1$ &  10 & $\approx 1.6$ \\
        TANCA      & Campinas, BR & $-22.8$ & $-47.1$ & 680 & $\approx 9.3$ \\
        Asunci\'on & Paraguay     & $-25.3$ & $-57.6$ & 126 & $\approx 10.7$ \\
        \hline
    \end{tabular}
\end{table}

\subsection{Marambio Station}
\label{S-marambio}
The Neurus WCD is installed at the Antarctic Space Weather Laboratory of LAMP (Laboratorio Argentino de Meteorología del esPacio), an interinstitutional group from Buenos Aires, Argentina \citep{LANABERE20202223}, dedicated to space weather research, instrumental development, R2O-O2R activities, atmospheric effects studies, and real-time monitoring of space weather conditions. The laboratory was installed during the 2019–2020 Argentine Antarctic campaign at the Argentine Marambio Base, located at $64.2^\circ$\,S, $56.7^\circ$\,W, and 198\,m a.s.l. The site is characterized by a vertical geomagnetic cutoff rigidity of $R_c \approx 2.1$\,GV. The Neurus WCD has been continuously operating since its installation \citep{Santos2023,Santos2025}.

\begin{figure}[h]
    \centering
    \includegraphics[width=0.55\textwidth]{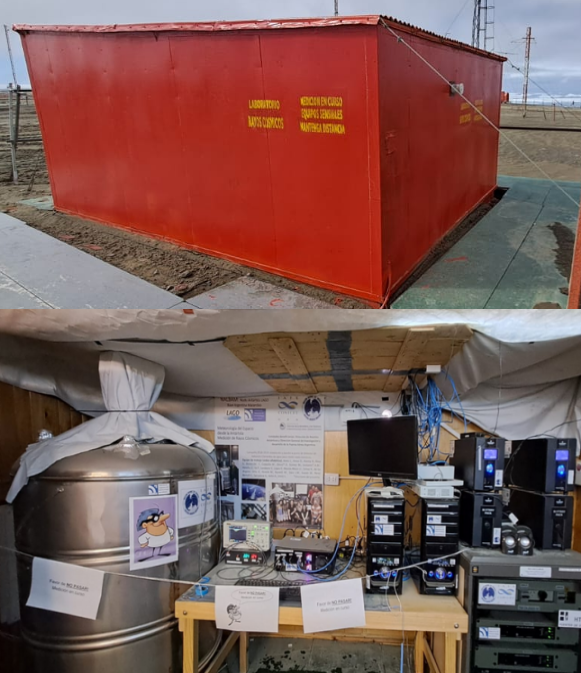}
    \caption{External view of the LAMP Antarctic Space Weather laboratory and the Neurus WCD installed at the Marambio Base. The figure shows the laboratory infrastructure and the detector setup deployed for cosmic-ray measurements.}
    \label{F-marambio}
\end{figure}

The WCD consists of three main components: a cylindrical tank, a photomultiplier tube (PMT), and the data acquisition system (DAQ) (see Figure \ref{F-marambio}). The tank is made of stainless steel, with a diameter of 0.96 m and a height of 1 m, and is filled with purified water. The detector setup also includes a GPS receiver for time synchronization, a meteorological station, and a telemetry system that provides 5-min real-time monitoring and data transmission to the LAMP servers in Buenos Aires.

The Neurus WCD DAQ was implemented in two configurations. In Mode A, a Rigol DS1102E oscilloscope in rate mode counts pulses exceeding a peak threshold of 0.02 V, and a communication system connected to a computer records the count rate. In Mode B, a Red Pitaya board runs an FPGA programmed with the LAGO acquisition system, enabling recording of the traces of all detected pulses \citep{Arnaldi2019}.

\subsection{San~Mart\'in Station}
\label{S-sanmartin}
\begin{figure}[h]
    \centering
    \includegraphics[width=0.8\textwidth]{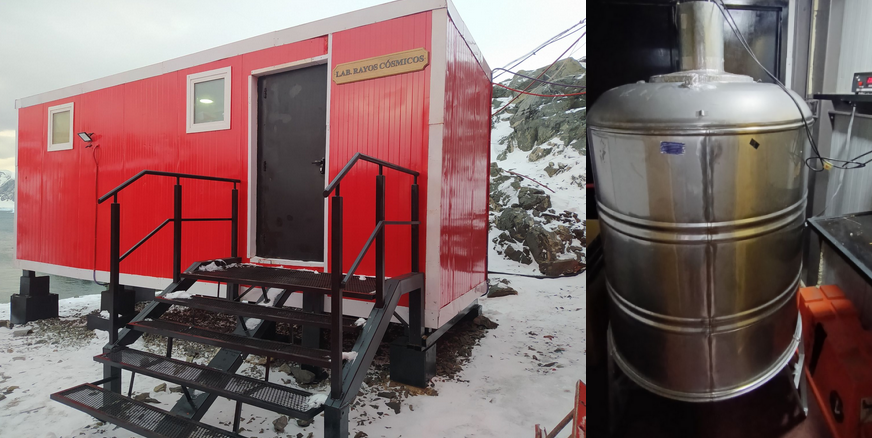}
    \caption{External view of the LAMP Antarctic Space Weather laboratory and the WCD installed at the San Martín Base. The figure shows the laboratory infrastructure and the detector setup deployed for cosmic-ray measurements.}
    \label{F-sanmartin}
\end{figure}

San Martín WCD is the second Antarctic station considered in this work and belongs to the same LAMP experimental program. It has the same instrumental configuration and data acquisition system as the Neurus WCD described in Section \ref{S-marambio}. We installed the detector in 2024 at the Argentine San Martín Antarctic Base, located at $68.1^\circ$\,S, $67.1^\circ$\,W, at sea level and ~600 km from the Marambio Base (see Figure \ref{F-sanmartin}). Since its installation, the San Martín WCD has operated continuously. Among all LAGO stations, this site has the lowest geomagnetic cutoff rigidity ($R_c \approx 1.6$\,GV), making it most sensitive to low-energy primary cosmic rays.

\subsection{TANCA Station}
\label{S-tanca}
TANCA is a WCD installed at the Instituto de F\'isica Gleb Wataghin, Universidade Estadual de Campinas (UNICAMP), Campinas, Brazil ($22.8^\circ$\,S, $47.1^\circ$\,W, 680\,m a.s.l.). Its position near the Tropic of Capricorn yields $R_c \approx 9.3$\,GV \citep{Tanca2025}, making it one of South America's high-$R_c$ LAGO nodes.

The DAQ system digitizes the PMT signals using a CAEN DT5720 waveform digitizer, operating at a sampling rate of $250\,\mathrm{MS\,s^{-1}}$ with 14-bit resolution. In parallel, an Arduino Uno microcontroller records local atmospheric conditions using a Bosch BMP280 barometric pressure sensor and auxiliary temperature and humidity sensors, providing the ancillary environmental data required for the barometric-coefficient correction applied to the muon count rate. A dedicated C/C++ acquisition routine merges the digitised PMT waveforms and the environmental readings into a single data stream, organised as ROOT \texttt{TTree} structures and stored in hourly ROOT-format files.

We equalised the relative gain of the three PMTs using the single-particle charge spectrum of each tube, obtained in auto-trigger mode with a discrimination threshold of $-15\,\mathrm{mV}$. Under this configuration, the charge spectrum shows a well-defined peak corresponding to single muons passing through the water volume, which we used as the calibration reference. We iteratively tuned the high voltage on each PMT until the muon peak converged to the same charge value in all three spectra, ensuring a uniform relative response across the PMTs.
\begin{figure}[h]
    \centering
    \includegraphics[width=0.55\textwidth]{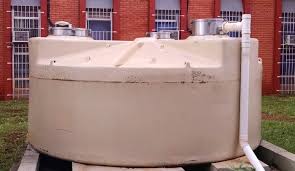}
    \caption{Photograph of the TANCA water-Cherenkov detector at UNICAMP. }
    \label{F-tanca}
\end{figure}

\subsection{Asunci\'on Station}
\label{S-asuncion}

The Asunción station is located at the Scientific Instrumentation Laboratory of the Faculty of Engineering, Universidad Nacional de Asunción, in San Lorenzo, Paraguay ($25.3^\circ$\,S, $57.6^\circ$\,W, 126\,m a.s.l.), where the vertical geomagnetic cutoff rigidity is approximately $R_c \approx 10.7$\,GV.

The detector consists of three $50 \times 50$ cm$^2$ Eljen EJ200 scintillator plates stacked vertically, separated a $5$ cm. Each scintillator is instrumented with four Hamamatsu S13360-6050VE silicon photomultipliers (SiPMs), whose outputs are summed by a summing amplifier, yielding one readout channel per scintillator; a custom DAQ processes the signals.

The DAQ comprises a signal-conditioning stage with individually adjustable thresholds for each channel and a digital-processing stage based on a Digilent Basys 3 FPGA. The FPGA implements coincidence logic to register single counts and double- and triple-coincidence events within a $100$ ns time window. A $1$ cm-thick steel plate between the middle and bottom scintillator plates further suppresses low-energy particle contributions to the coincidence channels. After acquisition, the acquisition software transfers the FPGA data and stores it in a MySQL database hosted on a local computer, following the implementation described by \citet{Molina2025}.
\begin{figure}
    \centering
    \includegraphics[width=0.55\textwidth]{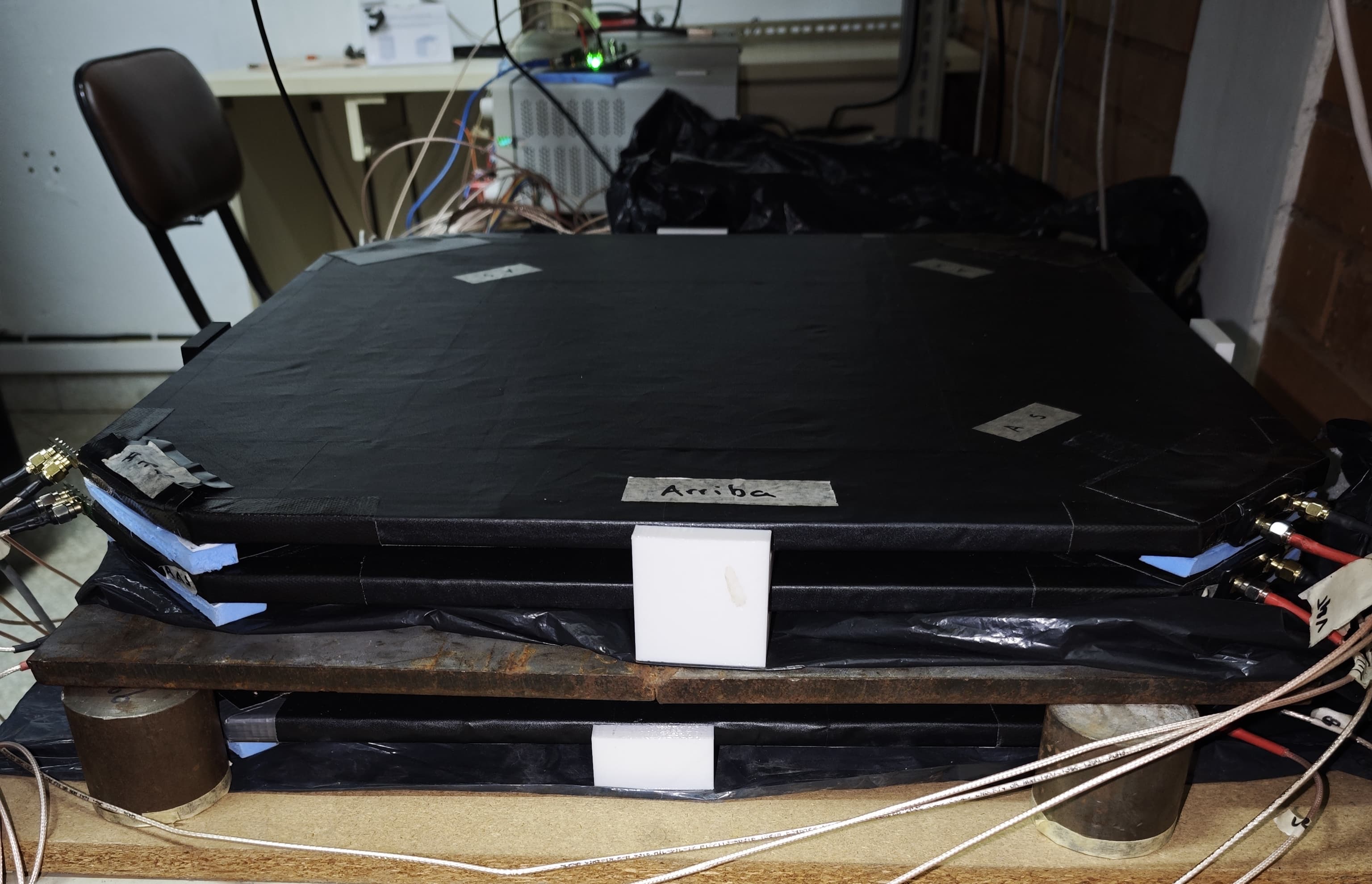}
    \caption{Photograph of the Asunción detector at FIUNA (adapted from \citealp{Molina2025}).}
    \label{F-asuncion}
\end{figure}

\section{Simulated Detector Response to Secondary Cosmic Rays}
\label{S-arti}
To investigate the LAGO WCDs' response to incident atmospheric radiation, the LAGO Collaboration developed the Atmospheric Ray Tracing and Interaction simulation framework (ARTI) \,\citep{SarmientocanoEtal2022}. Using Monte Carlo simulations, ARTI disentangles the WCD response into its secondary components, separating the muonic signal from electromagnetic and hadronic contributions. Consequently, we can design site-specific methodologies for each WCD to isolate the muon response, effectively turning the WCD into a ground-level muon counter sensitive to variations in the primary cosmic-ray flux\,\citep{asorey2018preliminary}.

The ARTI framework couples extensive air-shower simulations performed with CORSIKA\,\citep{heck1998corsika} to a Geant4-based detector model, while accounting for the local atmospheric and geomagnetic conditions at each site \citep{asorey2018preliminary}. In this work, we simulated detectors at Marambio, San Mart\'in, and TANCA. A site-specific atmospheric profile was constructed using data from the Global Data Assimilation System (GDAS), following \citet{MITRA2020102470}. For each location, we obtained the profile used in CORSIKA (v77402) by averaging the four atmospheric profiles available at 6-hour intervals throughout 1 January 2026, using the QGSJET-II and GHEISHA hadronic interaction models. Following the procedure described by \citet{SarmientocanoEtal2022}, we simulated an exposure equivalent to 24 hours of primary cosmic-ray flux for each site to obtain the expected distribution of secondary particles at ground level.

\begin{figure}[t]
    \centering
    \includegraphics[width=0.48\textwidth]{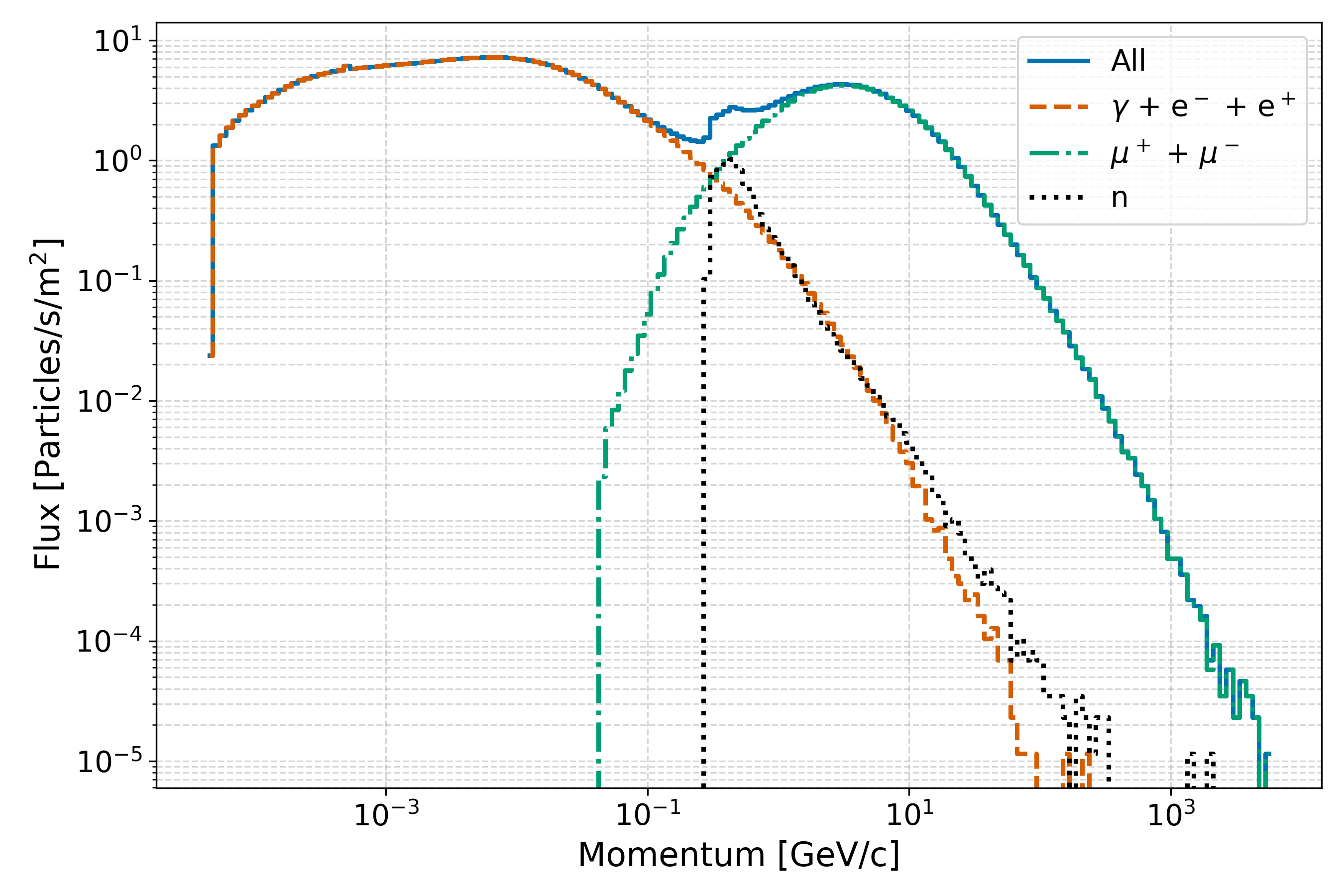}\hfill
    \includegraphics[width=0.48\textwidth]{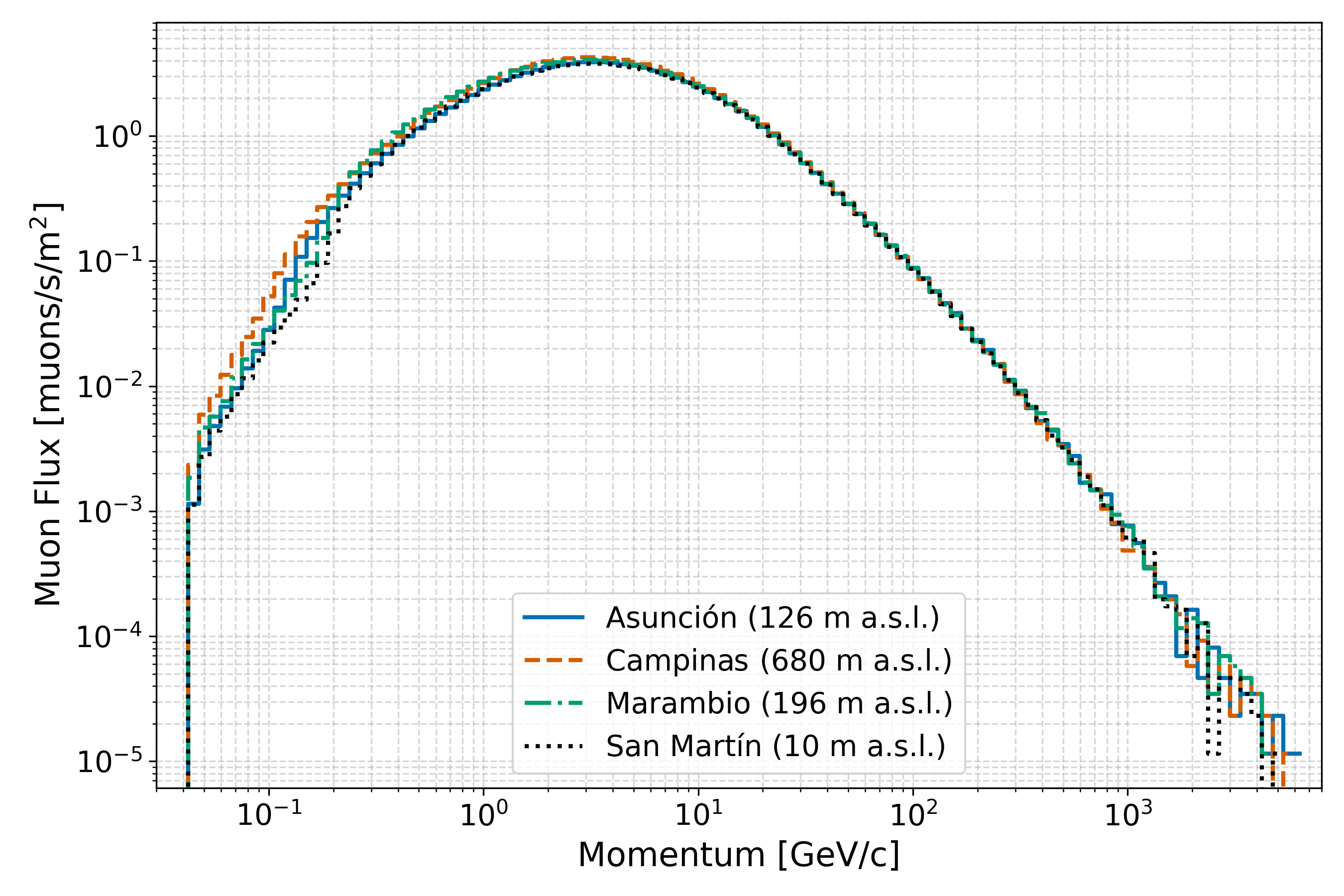}
    \vspace{0.3cm}
    \includegraphics[width=0.48\textwidth]{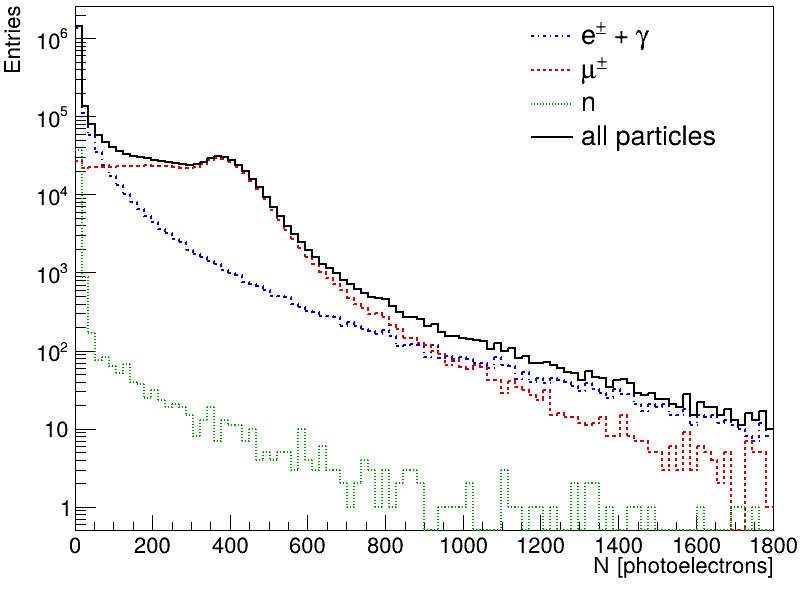}\hfill
    \includegraphics[width=0.48\textwidth]{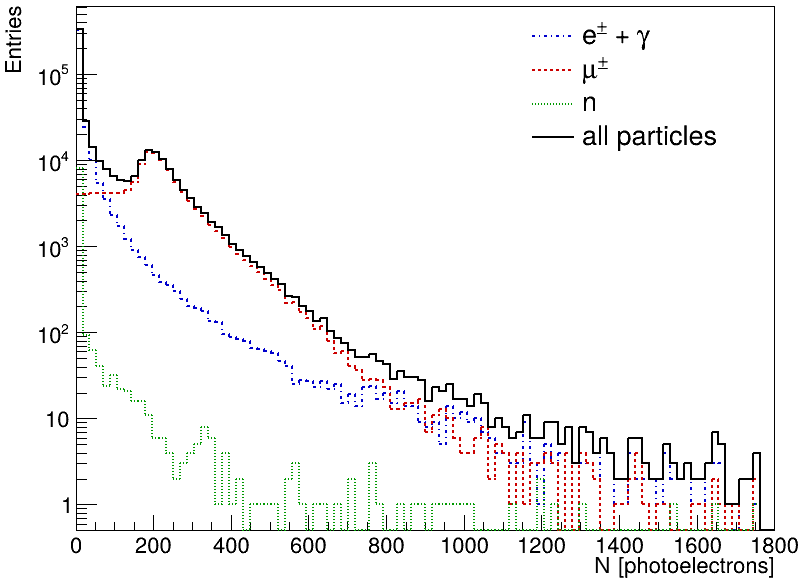}
    \caption{Simulated secondary particle distributions at ground level obtained with ARTI for the study sites \citep{SarmientocanoEtal2022}, alongside the corresponding WCD charge response. Top-left: Integrated secondary momentum spectra decomposed by particle species ($\gamma+e^\pm$, $\mu^\pm$, and $n$). Top-right: Ground-level momentum spectra for the muonic component across the four detector sites. Bottom panels: Simulated total integrated charge distributions collected by the WCD, highlighting the discrimination between the low-energy electromagnetic background and the characteristic vertical equivalent muon (VEM) peak for Marambio (on left) and TANCA (on right) stations. The horizontal axis gives the total number of photoelectrons, $N$, collected by the PMTs of each detector (one for Marambio, three for TANCA).}
    \label{F-arti}
\end{figure}

Figure\,\ref{F-arti} summarises the two-step simulation pipeline implemented in ARTI. The top-left panel illustrates the ground-level flux as a function of particle momentum for a representative site, clearly showing the dominance of the electromagnetic component ($\gamma + e^{\pm}$) below $10^{-1}$\,GeV/c, whereas muons ($\mu^{\pm}$) dominate the high-momentum regime ($>1$\,GeV/c), with a minor contribution from neutrons, $n$. The top-right panel shows the variation of the vertical muon flux across the four monitoring stations, reflecting the interplay among atmospheric profiles, local geomagnetic cutoffs, and altitudes ranging from 10\,m a.s.l.\ (San Mart\'in) to 680\,m a.s.l.\ (Campinas/TANCA).

The lower panels display the resulting WCD response functions in terms of the total deposited charge obtained after propagating these secondary spectra through the Geant4 detector model. The distributions show a clear structural separation: a high-amplitude soft component at low photoelectron counts dominated by electromagnetic showers, followed by the characteristic hump produced by relativistic muons traversing the water volume (Vertical Equivalent Muon, VEM), located at $\sim 350$--$400$\,photoelectrons for Marambio and at $\sim 200$\,photoelectrons for TANCA. The two panels illustrate the two WCD geometries present in the network: the small single-PMT tanks of the Antarctic stations (Marambio and San~Mart\'in share the same design) and the large three-PMT tank of TANCA (Table~\ref{T-arti-sites}). The VEM peak position depends on the tank geometry and PMT configuration, but in both cases the muonic hump is clearly separated from the electromagnetic component. This prominent muonic signature demonstrates that an appropriate charge threshold or multi-PMT coincidence criteria can effectively isolate the muon contribution from the background noise and soft secondary radiation.

The simulation parameters are described in Table \ref{T-arti-sites}, while the altitude a.s.l.\ and rigidity cutoff values match those in Table \ref{T-detectors}.

\begin{table}[t]
    \centering
    \caption{Site-specific parameters used in the ARTI and WCD simulations.
    The geomagnetic field components were obtained from the IGRF model~\protect\citep{IGRFcalc}.}
    
    \label{T-arti-sites}

    \begin{tabular}{lcccc}
        \hline
        \textbf{Site} &
        $\boldsymbol{B_{\mathrm{N}}}$ &
        $\boldsymbol{B_{\mathrm{Z}}}$ &
        \textbf{WCD diameter} &
        \textbf{Water height} \\
        &
        \textbf{(IGRF, $\boldsymbol{\mu}$T)} &
        \textbf{(IGRF, $\boldsymbol{\mu}$T)} &
        \textbf{(m)} &
        \textbf{(m)} \\
        \hline
        Marambio     & 19.17 & -29.97 & 0.96 & 1.0 \\
        San Mart\'in & 19.24 & -34.79 & 0.96 & 1.0 \\
        TANCA        & 16.31 & -14.54 & 3.68 & 1.14 \\
        \hline
    \end{tabular}
\end{table}

\section{Results and Discussion}
\label{S-results}

\subsection{LAGO and Mexico neutron monitor count-rate time series}
\begin{figure}
	\centering
	\includegraphics[width=\textwidth]{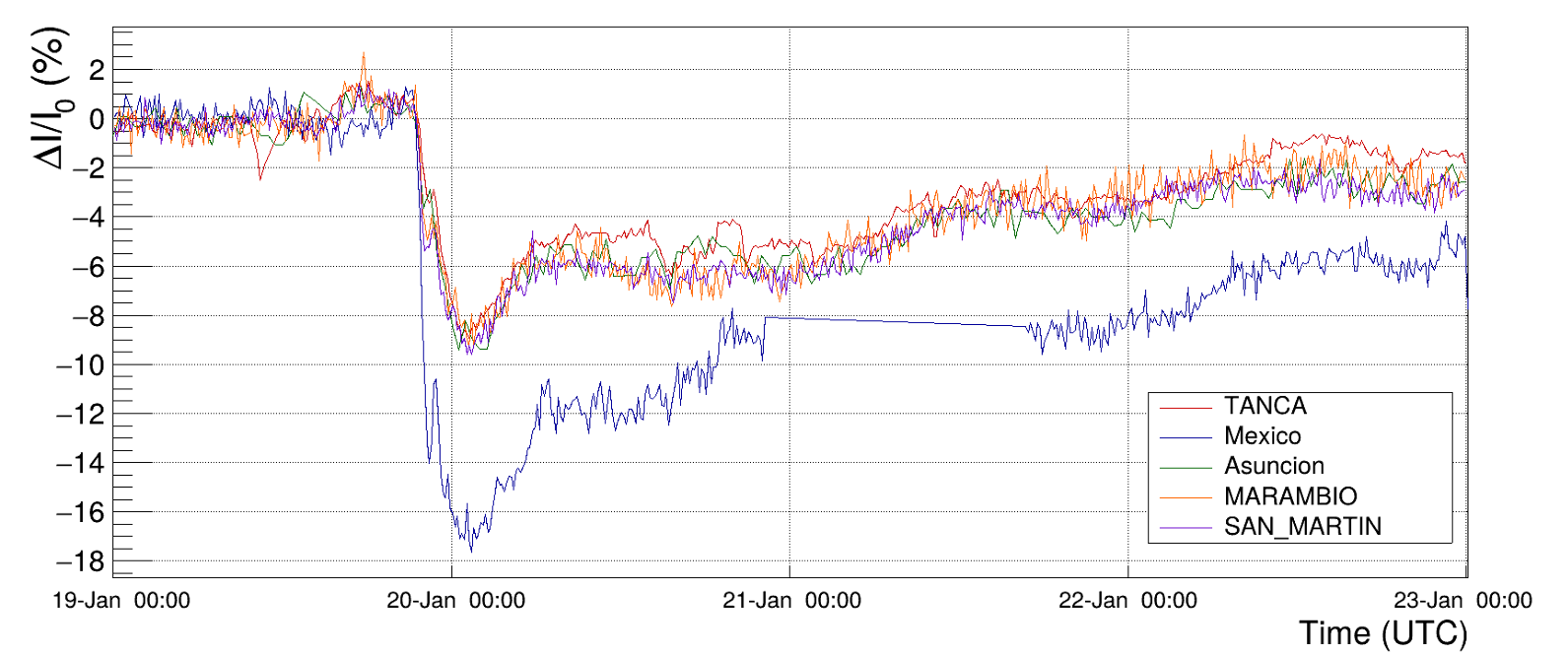}
	\caption{Multi-detector comparison of the Forbush decrease, pressure-corrected count-rate time series for the four LAGO stations and the Mexico neutron monitor, expressed as percent deviation from the pre-event baseline.}
	\label{F-nm}
\end{figure}

Figure~\ref{F-nm} shows the pressure-corrected count-rate time series, expressed as percent deviation from the pre-event baseline ($\Delta I/I_0$), for the four LAGO stations (TANCA, Asunci\'on, Marambio, and San Mart\'in) and the Mexico Neutron Monitor ($R_c = 8.2$~GV; \citealp{NMDB2025}) at 10-minute resolution over the interval 19--23 January 2026.

All five detectors register a nearly simultaneous onset of the Forbush decrease at approximately 21:40~UT on 19 January. The four LAGO stations reach comparable minimum depressions of $9.09$--$9.41\%$ between 00:30 and 01:15~UT on 20 January, while the Mexico neutron monitor exhibits a substantially deeper minimum of about $17\%$ around the same time.
The FD amplitudes across stations should not be interpreted only in terms of $R_c$. The asymptotic directions of the primaries differ from site to site; for example, primaries with $R \approx 5$--$10$~GV detected at Marambio arrive from asymptotic directions over Africa \citep{Santos2023}. In addition, the median rigidity of the muonic response is well above the local cutoff, which reduces the sensitivity of the WCD count rate to $R_c$.

The similarity of the LAGO amplitudes should therefore not be read as evidence of a weak rigidity dependence of the modulation. For the WCDs, whose count rate includes a large muonic contribution, the median rigidity of the detected primaries lies well above the local cutoff, so that the vertical $R_c$ is a poor proxy for the rigidity range actually sampled. In additTable~\ref{T-detectors} are quiet-time values; their reduction during the G4 we have not quantified storm, expected to be largest at the high-$R_c$ stations, has not been quantified here. A rigidity-resolved interpretation requires storm-time cutoffs and asymptotic directions, e.g.\ from trajectory calculations with the OTSO code \citep{Larsen2023OTSO}, which we leave for future work. The deeper decrease at the Mexico neutron monitor is consistent with the nucleonic component's lower effective energy threshold, which makes neutron monitors more responsive to the softening of the primary galactic-cosmic-ray spectrum during the ICME passage.

\subsection{Magnetic-field and plasma context}
\label{S-plasma}

\begin{figure*}[th]
    \centering
    \includegraphics[width=\textwidth]{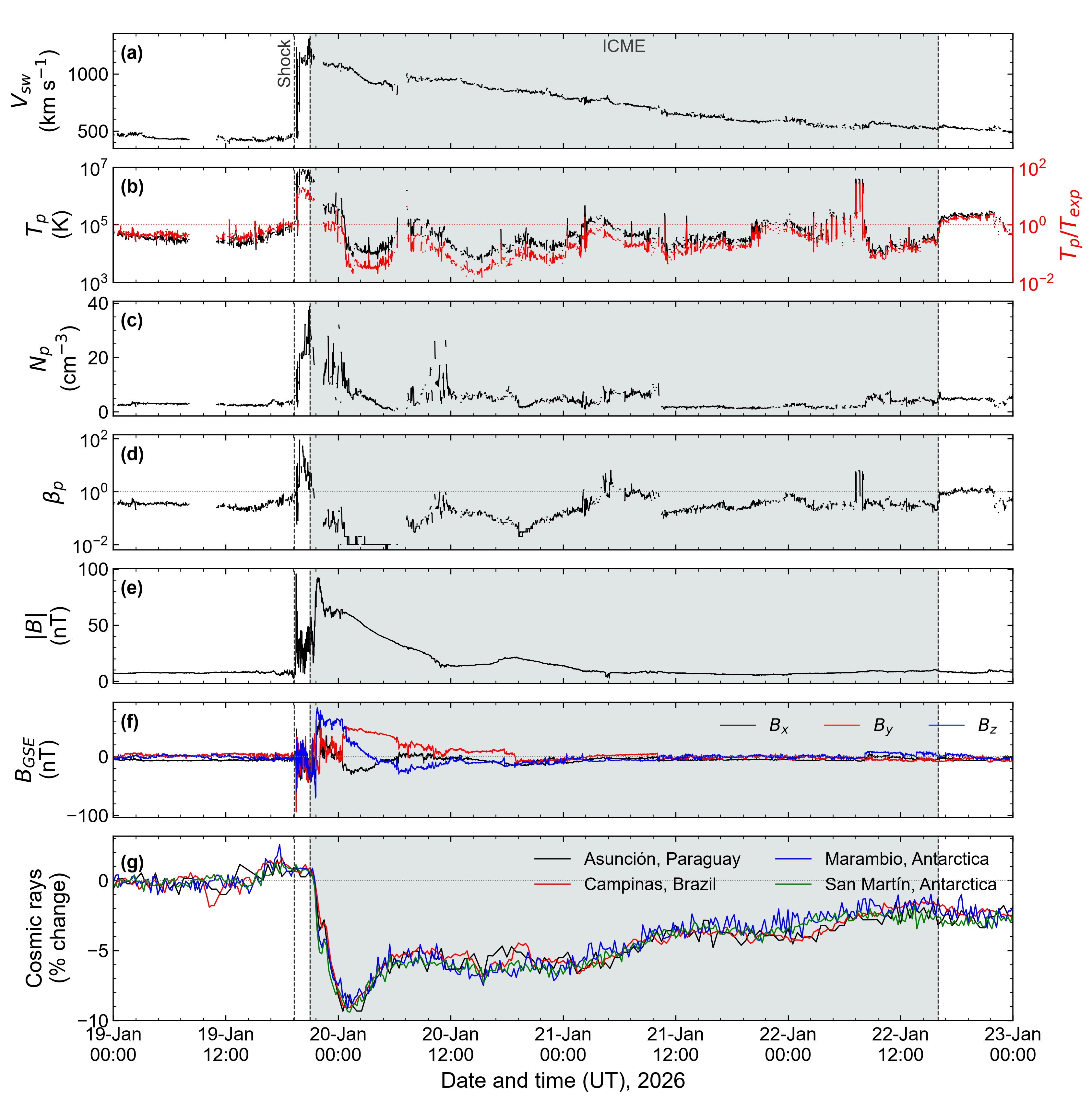}
    \caption{Interplanetary conditions and ground-level cosmic-ray response associated with the ICME observed on 19--22 January 2026. Panels (a)--(d) show the solar-wind speed, proton temperature (black, left axis) together with the $T_p/T_{\mathrm{exp}}$ ratio (red, right axis), proton number density, and proton plasma beta, respectively. Panels (e) and (f) show the IMF magnitude and its GSE components. Panel (g) presents the relative cosmic-ray intensity variations measured at Asunci\'on, Campinas, Marambio, and San Mart\'in. The first vertical dashed line marks the interplanetary shock at 19:17~UT on 19 January. The interval between the shock and the ICME leading edge corresponds to the compressed sheath-like region discussed in the text. The shaded region delimited by the second and third dashed lines indicates the ICME interval, from 21:00~UT on 19 January to 16:00~UT on 22 January.}
    \label{fig:icme_overview}
\end{figure*}

Figure~\ref{fig:icme_overview} summarizes the near-Earth interplanetary conditions during the passage of the 19 January 2026 ICME and compares them with the ground-level cosmic-ray response recorded by the four LAGO stations. The solar-wind speed ($V_{\mathrm{sw}}$), proton temperature ($T_p$), proton number density ($N_p$), and proton plasma beta ($\beta_p$), shown in panels (a)--(d), were obtained from the 1-min high-resolution OMNI data set distributed by the NASA/GSFC Space Physics Data Facility through OMNIWeb \citep{OMNIWEB}. Panel (b) also shows the ratio $T_p/T_{\mathrm{exp}}$, where $T_{\mathrm{exp}}$ is the expected proton temperature calculated from $V_{\mathrm{sw}}$ using the Richardson--Cane empirical relation \citep{RichardsonCane1995}. The horizontal dotted line indicates $T_p/T_{\mathrm{exp}}=1$.

The magnitude and vector components of the interplanetary magnetic field (IMF), shown in panels (e) and (f), respectively, were obtained from the 16-s Level-2 measurements of the Magnetic Field Experiment (MAG) aboard the Advanced Composition Explorer (ACE), available from the ACE Science Center \citep{ACEMAG}. The IMF components are presented in Geocentric Solar Ecliptic (GSE) coordinates. ACE magnetic-field measurements were used because they provide substantially better temporal coverage than the corresponding high-resolution OMNI magnetic-field data during this event.  Comparison with the near-Earth OMNI magnetic-field measurements indicates a variable propagation delay between ACE and Earth; therefore, ACE data are used primarily to characterize the magnetic structure of the disturbance, whereas the shock time and ICME boundaries used for comparison with the LAGO observations are referenced to near-Earth data products and the Richardson and Cane ICME catalog \citep{RichardsonCane2010}.

The interplanetary disturbance is marked by a shock arrival at 19:17~UT on 19 January, followed by abrupt enhancements in the solar wind speed, proton temperature, proton density and magnetic-field magnitude. The components of the magnetic-field become highly variable, indicating a compressed and disturbed sheath-like region downstream of the shock.  These signatures characterise the first strongly perturbed region the near-Earth environment encounters.

Panel (g) presents the relative cosmic-ray intensity variations recorded at Asunci\'on, Campinas, Marambio, and San Mart\'in. The four detectors show a nearly simultaneous Forbush decrease, with an onset at approximately 21:40~UT, about 2.4~h after the shock arrived. The temporal ordering suggests that the shock discontinuity itself was not a region associated with the cosmic-ray decrease.

A noticeable change in the slope of the cosmic-ray decrease occurs during the transition from the strongly fluctuating and compressed magnetic-field configuration following the shock toward a more organized magnetic field. This change is quickly followed by a second deepening toward the minimum. The feature is observed simultaneously at all four LAGO stations. Its temporal correspondence with the evolving magnetic-field configuration is consistent with sequential modulation by different ICME regions.

The count rates continue to decrease after the FD onset and reach their minimum between 00:30 and 01:15~UT on 20 January, while the magnetic field remains strongly enhanced and the plasma parameters evolve toward the low-$\beta_p$, low-$T_p/T_{\mathrm{exp}}$ conditions characteristic of the ICME ejecta. The maximum depression reaches $9.34\%$ at Asunci\'on, $9.09\%$ at Campinas, $9.15\%$ at Marambio, and $9.41\%$ at San Mart\'in. A gradual recovery follows throughout the ICME passage.

This sequence is consistent with a modulation scenario in which the compressed, highly disturbed plasma behind the shock drives the initial decrease, while the subsequent ejecta further deepens the minimum. The four LAGO stations reproduce this temporal evolution nearly simultaneously, supporting an interplanetary origin for the observed decrease structure.

After the minimum, the count rates begin a gradual recovery. This recovery develops while the magnetic-field strength and the overall solar-wind disturbance progressively relax during the ICME passage.

\subsection{CME Transit Time and the Timing of the Forbush Decrease}
\label{S-dbm}

We reconstructed the CME's early kinematics using the Graduated Cylindrical Shell (GCS) model \citep{2006Apj} applied to multi-viewpoint SOHO/LASCO and STEREO-A coronagraph observations. Assuming self-similar expansion, GCS provides the heliocentric distance $r(t)$ to the leading edge of the flux-rope front at each time step. In the low corona ($r \lesssim 20\,R_\odot$), where the Lorentz force still dominates, the height--time profile is commonly represented by a second-order polynomial \citep{Bein2011},
\begin{equation}
r(t) = r_0 + v_0 t + \frac{1}{2} a_0 t^2 ,
\label{equation1}
\end{equation}
where $r_0$ and $v_0$ are the height and speed at the reference time and $a_0$ is the acceleration. The fit gives $v_0 = 1736 \pm 109$~km\,s$^{-1}$ at $r_0 = 7\,R_\odot$ (18:30~UT on 18 January). We used the last GCS point ($77\,R_\odot$, 01:30~UT on 19 January) as the starting point for interplanetary propagation.

Beyond the corona, the CME dynamics are governed by the aerodynamic drag exerted by the ambient solar wind \citep{Cargill1996,Cargill2004,VrsnakGopalswamy2002}. In the Drag-Based Model \citep[DBM;][]{Vrsnak2013} the drag acceleration is
\begin{equation}
a_D = -\gamma\,(v - w)\,|v - w| , \qquad
\gamma = \frac{c_d\,A\,\rho_{sw}}{M} ,
\label{equation2}
\end{equation}
where $v$ is the CME speed, $w$ the solar-wind speed, $c_d \simeq 1$ the drag coefficient, $A$ the CME cross-section, $\rho_{sw}$ the ambient density, and $M$ the CME mass. For constant $\gamma$ and $w$ and $v_0 > w$, the solution is
\begin{equation}
\begin{gathered}
v(t) = w + \frac{v_0 - w}{1 + \gamma (v_0 - w)\, t} , \\
r(t) = r_0 + w t + \frac{1}{\gamma}\ln\left[1 + \gamma (v_0 - w)\, t\right] ,
\end{gathered}
\label{equation3}
\end{equation}
and the arrival time follows from $r(t_{1\,\mathrm{AU}}) = 1$~AU. Because $\gamma \propto \rho_{sw}$, a medium depleted by preceding eruptions lowers the drag (preconditioning; \citealp{Liu2014,TemmerNitta2015,Desai2020}). For the extreme event of 23 July 2012, \citet{Liu2014} required $\gamma \sim 10^{-8}$~km$^{-1}$, about one order of magnitude below the typical value of $10^{-7}$~km$^{-1}$ \citep{Vrsnak2013}.

\begin{figure}
\centering
\IfFileExists{figures/propagacion_DBM_comparacion.png}{%
  \includegraphics[width=0.85\textwidth]{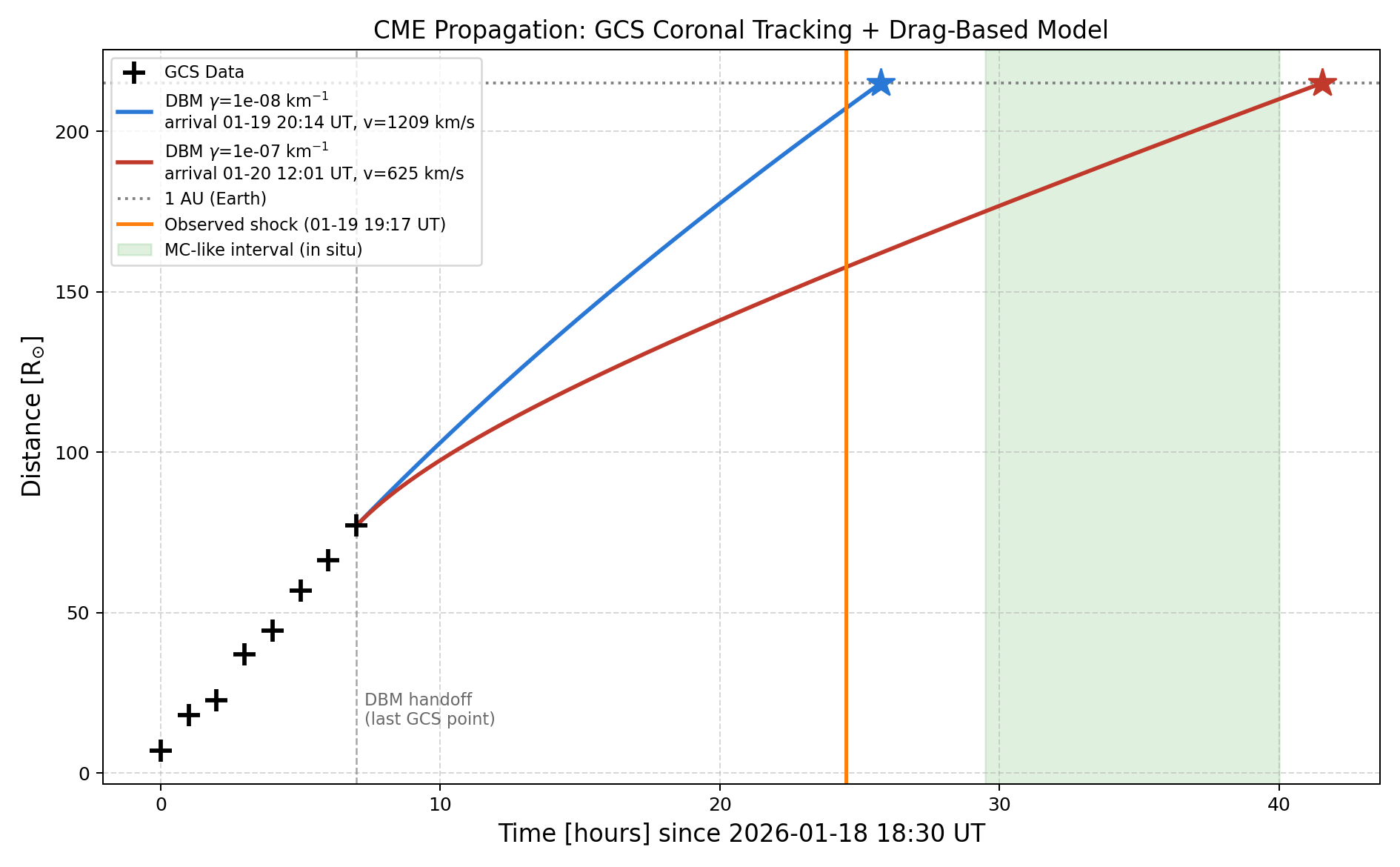}}{%
  \fbox{\parbox[c][5cm][c]{0.8\textwidth}{\centering FIGURE MISSING: figures/gcs\_dbm\_heighttime.png}}}
\caption{Height--time profile of the CME leading edge from GCS (black crosses) and DBM propagation for $\gamma = 10^{-8}$ and $10^{-7}$~km$^{-1}$ (colored curves). The orange vertical line marks the shock arrival at L1 (19:17~UT, 19 January); the shaded band marks the magnetic-cloud-like interval.}
\label{F-dbm}
\end{figure}

We adopt $w = 550$~km\,s$^{-1}$, the pre-event value measured by WIND, and compare these two regimes (Table~\ref{T-dbm}, Figure~\ref{F-dbm}). We use the one-dimensional DBM with constant $\gamma$ and $w$ as a consistency test rather than as a forecast: its two-parameter analytic solution isolates the role of the drag regime. More elaborate variants, with distance-dependent $\gamma(r)$ and $w(r)$, parameters fitted to the heliospheric observations, and a cone geometry for the CME front \citep{Zic2015}, require inputs that are poorly constrained in a heliosphere disturbed by several preceding CMEs.

\begin{table}
\caption{DBM predictions for the arrival at 1~AU of the CME front tracked by GCS, for $w = 550$~km\,s$^{-1}$.}
\label{T-dbm}
\begin{tabular}{ccc}
\hline
$\gamma$ [km$^{-1}$] & Predicted arrival (UT) & $v_{1\,\mathrm{AU}}$ [km\,s$^{-1}$] \\
\hline
$1\times10^{-8}$ & 2026-01-19 20:14 & 1209 \\
$1\times10^{-7}$ & 2026-01-20 12:01 &  625 \\
\hline
\end{tabular}
\end{table}

When comparing these predictions with the \textit{in-situ} data, keep in mind that GCS tracks the leading edge of the flux-rope front, not the shock driven ahead of it. Therefore, compare the DBM arrival with the ICME leading edge rather than the shock. With low drag, the predicted arrival (20:14~UT on 19 January) falls between the shock crossing of L1 (19:17~UT) and the ICME leading edge (21:00~UT, Section~\ref{S-plasma}). It also precedes by about 1.4~h the onset of the FD at the four LAGO stations ($\approx$\,21:40~UT). With typical drag, the front would arrive at 12:01~UT on 20 January, about 11~h after the FD minimum had already been reached (00:30\,--\,01:15~UT). The FD timing therefore independently supports the low-drag scenario. It also confirms that the sheath and ejecta of this CME, not a later structure, produce the decrease.

The low-drag regime is plausible given the preceding days' eruptive activity. Ten CMEs were launched between 15 and 18 January. We propagated each with the DBM from its GCS parameters (longitude, latitude, aspect ratio, and half-angle) and checked whether the reconstructed volumes overlapped in three dimensions. Although the one-dimensional apex trajectories of the 18 January CME cross those of most earlier events, only one of them (CME~7, launched on 17 January) overlaps it geometrically before 1~AU, at $\approx 138\,R_\odot$ around 07:00~UT on 19 January. The other events were propagating in different directions. The material swept up by these preceding ejecta is consistent with the reduced drag required to reproduce the observed transit.

\section{Conclusions}
\label{S-conclusions}

We present a multi-station analysis of the FD on 19--20 January 2026, combining ground-level cosmic-ray observations from the LAGO network with the solar, interplanetary, and geomagnetic context of the driving event. The main findings are:

\begin{enumerate}
  \item The FD was driven by a full-halo CME associated with an X1.95 flare from AR~4341 on 18 January (soft X-ray peak at 18:09~UT). GCS reconstruction yields an initial speed $v_0 = 1736 \pm 109$~km\,s$^{-1}$. DBM propagation with a low drag parameter ($\gamma = 10^{-8}$~km$^{-1}$) places the CME front at 1~AU at 20:14~UT on 19 January. This lies between the shock arrival at L1 (19:17~UT) and the ICME leading edge (21:00~UT), and about 1.4~h before the FD onset. A typical drag parameter ($\gamma = 10^{-7}$~km$^{-1}$) would delay the arrival until 12:01~UT on 20 January, after the FD minimum. The $\approx 25$-h transit therefore requires a medium preconditioned by preceding CMEs. Of the ten CMEs launched between 15 and 18 January, only one (CME~7) overlapped geometrically with the 18 January CME before it reached Earth.
  \item The interplanetary shock (19:17~UT, 19 January) and subsequent G4 geomagnetic storm produced a nearly simultaneous FD onset at all four LAGO stations ($\sim$21:40~UT, 19 January), with comparable  maximum depressions of $9.09$--$9.41\%$ at Marambio, San~Mart\'in, TANCA, and Asunci\'on, and minima reached between 00:30 and  01:15~UT on 20 January.
  \item Despite the different local cutoff rigidities of the four stations (1.6--10.7~GV), their FD amplitudes are very similar. Because the median rigidity of the WCD response lies well above the local cutoff, and we did not quantify storm-time changes in cutoff rigidity and asymptotic directions, this similarity alone does not constrain the rigidity dependence of the modulation. The Mexico neutron monitor shows a substantially deeper ($\sim$17\%) and more slowly recovering decrease, consistent with nucleonic- and muon-sensitive detectors responding differently to the ICME-modulated galactic cosmic-ray spectrum.
  \item ARTI/Geant4 simulations of the WCD responses at each site reproduce the expected separation between the electromagnetic background and the muonic (VEM) signal, with the characteristic muon peak, supporting the use of charge-based criteria to isolate the muonic contribution to the count rate.
\end{enumerate}

Taken together, these results demonstrate the LAGO network's ability to detect and characterise Forbush decreases across a wide range of geomagnetic rigidity cutoffs and detector technologies, and to link ground-level observations to upstream solar and interplanetary drivers through combined GCS/DBM modelling. The comparison with the Mexico neutron monitor underscores the complementary role of nucleonic- and muonic-component detectors in characterising the rigidity dependence of the cosmic-ray modulation. Future work will extend this analysis to a resolved determination of the modulation spectral index and to a systematic comparison with additional NMDB/GMDN stations.

\begin{acks}
We acknowledge the NMDB database (\url{www.nmdb.eu}), founded under the European Union's FP7 program (contract no.\ 213007), for providing data. We acknowledge the use of NASA/GSFC's Space Physics Data Facility's OMNIWeb service and OMNI data, and the ACE Science Center for the ACE/MAG data. During the preparation of this manuscript, the authors used an AI-based language tool (ChatGPT, OpenAI) to improve the language and readability of the text. After using this tool, the authors reviewed and edited the content as needed and take full responsibility for the content of the publication.
\end{acks}


\begin{fundinginformation}
S.D. and A.M.G. are members of the Carrera del Investigador Científico, CONICET. This work was supported by the Argentine grants PIP 11220220100345CO (CONICET) and UBACyT 20020190100247BA and PIDAE (UBA), the National Council for Scientific and Technological Development (CNPq), Brazil, under Grants 446365/2014-9 and 310473/2025-0, and São Paulo Research Foundation (FAPESP) under grants 19/10151-2 and 99/05404-3. The authors also acknowledge co-funding from the Programa Iberoamericano de Ciencia y Tecnología para el Desarrollo, CYTED, through the LAGO-INDICA network (Project 524RT0159-LAGO-INDICA: Infraestructura digital de ciencia abierta).
\end{fundinginformation}

\begin{authorcontribution}
N.A.S. contributed to the experimental description and revised the manuscript.
A.C.F. operated the TANCA detector, analysed its data, and compared the Forbush
decrease observed at TANCA with that recorded by the other detectors.
S.D. contributed to the experimental description and was responsible for the
operation and data analysis of the Antarctic water-Cherenkov detectors
(Marambio and San Mart\'{\i}n).
A.M.G. analysed the interplanetary medium and the chain of events from the Sun
to Earth.
J.M. and G.B.S.G. provided the experimental data of the Asunci\'on detector and analysed
the Forbush decrease observed there.
J.N. performed simulations of coronal mass ejections and analysed them.
L.A.N. and Y.D.B. analysed the interplanetary conditions and ground-level cosmic-ray response.
J.F.S.C. performed the Monte Carlo simulations with ARTI--CORSIKA.
M.S.D. and C.S.-C. performed Monte Carlo simulations with the ARTI--Meiga framework, including
the study of extensive air showers and of the water-Cherenkov detector response.
All authors reviewed and approved the final manuscript.
\end{authorcontribution}

\begin{ethics}
\begin{conflict}
The authors declare no conflicts of interest.
\end{conflict}
\end{ethics}

\begin{dataavailability}
The pressure-corrected 10-min count-rate time series of the four LAGO stations used in this study are openly available in the Unicamp Research Data Repository \citep{dataFD} under a CC BY 4.0 license. NMDB data are publicly available at \url{https://www.nmdb.eu}; OMNI and ACE data are available from NASA/GSFC SPDF OMNIWeb and the ACE Science Center, respectively.
\end{dataavailability}




\end{document}